\documentclass[reprint,amsmath,amssymb,aps,preprintnumbers,nofootinbib]{revtex4-2}

\usepackage[bookmarks=true, pdfpagelabels=true, colorlinks=true, linkcolor=black, urlcolor=black, citecolor=blue, anchorcolor=blue]{hyperref}
\usepackage[T1]{fontenc} 
\usepackage[dvipsnames]{xcolor}
\usepackage{slashed}
\usepackage{enumitem}
\usepackage{amsmath}
\usepackage{comment}
\usepackage[normalem]{ulem}
\usepackage{amssymb}
\usepackage{graphicx}
\usepackage[switch]{lineno} 

\newcommand{\mgFull}{{\sc MadGraph5\_aMC@NLO}~}

\newcommand{\C}[1]{\textcolor{black}{#1}}
\newcommand{\ie} {{\it i.e.}\;}

\begin{document}


\title{Signatures of leptophilic t-channel dark matter from active galactic nuclei}

\preprint{CP3-22-05}


\author{Marina Cermeño}
\email{marina.cermeno@uclouvain.be}
\author{Céline Degrande}%
 \email{celine.degrande@uclouvain.be}
\affiliation{%
 Centre for Cosmology, Particle Physics and Phenomenology (CP3),
Universit\'e Catholique de Louvain,
B-1348 Louvain-la-Neuve, Belgium
}%

\author{Luca Mantani}
\email{luca.mantani@maths.cam.ac.uk}
\affiliation{DAMTP, University of Cambridge, Wilberforce Road, Cambridge, CB3 0WA, United Kingdom}

\begin{abstract}

In this work, we study indirect photon signatures of leptophilic dark matter (DM) coming from Centaurus A, where a DM density spike is believed to have survived to date contrary to the case of our galaxy. We consider a model where DM is a Majorana fermion which interacts with right-handed electrons via a scalar mediator.
Assuming that the photons measured from the core of Cen A are coming from SM processes, we derive constraints \C{on the average annihilation cross section which are 7 orders of magnitude stronger than the ones from measurements of the Galactic Center. Focusing on the allowed parameter space range,} we calculate the flux of photons coming from the radiative DM-electron scattering in the AGN jets and the circular polarisation asymmetry of these photons. We find that this flux is two orders of magnitude lower than the background but its circular polarisation asymmetry can reach values close to $100\%$, indicating the need to experimentally exploit the high fraction of circular polarisation in order to detect these interactions.
Since the origin of the photons in the GeV-TeV range from Cen A is not completely clear and an exotic origin is compatible with the measurements as well, we also consider the scenario in which synchrotron radiation can only partially explain the photon flux and we fit the excess with signals coming from DM annihilation, finding a best fit for a DM candidate with a mass $m_{\tilde{\chi}}= 123$ GeV and a coupling $a_R = 0.018$.
\end{abstract}


\maketitle

\section{Introduction}

One of the most important challenges in modern cosmology and particle physics is the understanding of the nature of dark matter (DM). Despite the precise determination of the DM abundance in the presently accepted model for our Universe ($\Omega_{CDM} h^2= 0.120 \pm 0.001$) by the Planck collaboration~\cite{Planck:2020Id0}, its true identity remains unknown. Current theoretical efforts are aimed at finding
extensions of the Standard Model (SM) of particle physics, introducing new particles that would explain the astrophysical and cosmological observations. Likewise, experimental techniques are trying to detect these particles directly or indirectly, pushing further their sensitivities.

Indirect detection experiments look for an excess of SM products coming from DM interactions in high DM density regions of the sky.
In particular, conventional indirect photon searches are designed to attempt to measure an excess of photons coming from DM interactions, usually self-annihilations into SM particles, over the astrophysical background.

The center of galaxies are expected to host a huge amount of dark and baryonic matter. The inner region of our galaxy has been broadly explored in the context of DM searches, \C{however, younger galaxies with active galactic nuclei (AGN), such as Centaurus A (Cen A) or Messier 87 (M87), are expected to gather a higher DM component.} Therefore, the study of DM in their interior can help us to better characterize the properties of these unknown particles. In particular, Cen A and M87 are believed to possess a high DM density spike due to DM accretion onto supermassive black holes (BH) in the interior of these objects~\cite{Gondolo:1999ef}, which would have survived to date, contrary to the case of the Milky Way. %
Exploiting this fact, some DM models have been constrained. For example, constraints on the velocity-independent (s-wave) cross-section for the annihilation of weakly interacting massive particles (WIMPs) in the core of M87 have been derived in Refs.~\cite{Lacroix:2015lxa, Lacroix:2016qpq}, suggesting that thermal DM with masses $\lesssim 10^5$ GeV would be ruled out for this class of models. 
\C{However, if the constraint of the relic density is relaxed, this model is able to explain the anomalous peak of the photon flux spectrum of Cen A measured by Fermi-LAT between $2.5$ GeV and $5$ TeV with self-annihilations of non-thermal DM with masses of hundreds of GeV or some TeVs~\cite{Brown:2016sbl}.}

Moreover, these objects are well-known sources of high-energy particles, such as electrons and protons, which will interact with DM in the AGN jets. Therefore, we will not only expect signals from the self-annihilation of DM but also from its scattering with the baryonic and leptonic matter of the AGN jets. In particular, the total flux of photons coming from interactions between neutralinos and electrons in the AGN of Cen A is found to be high enough to be observable by Fermi LAT~\cite{Gorchtein:2010xa, Huang:2011dg, Gomez:2013qra}. However, the authors assume a jet power in electrons which is incompatible with the measurements as the jet model considered would provide a photon flux orders of magnitude higher than the observed one. Consequently, we re-evaluated the intensity of this kind of signatures. 

Furthermore, the photons produced in DM scattering with cosmic rays are expected to be circularly polarised, giving possibly an extra handle to distinguish them from the  background. 
Recent works~\cite{Kumar:2016cum, Elagin:2017cgu, Gorbunov:2016zxf, Boehm:2017nrl, Huang:2019ikw, Cermeno:2021rtk} have shown that DM interactions with SM particles can generate circular polarised signals in X rays or gamma rays.
A net circular polarisation can be observed in the sky when there is an excess of one photon polarization state over the other. As the notion of circular polarisation is related to parity (P) violation
due to the fact that photons flip helicity under parity, P must be violated in at least one of the dominant photon
emission processes. But P violation is not the only condition required, there must be either an asymmetry in the
number density of one of the particles in the initial state or CP must be violated as well. 
Accordingly, a P violating interaction of DM with electrons in a region of the sky where there are more electrons than positrons will be a source of circularly polarised photons \cite{Boehm:2017nrl, Cermeno:2021rtk}. 

Even if nowadays there are no experiment able to measure circular polarisation in gamma rays, this feature is important to understand the DM nature and could be exploited to detect DM.
Motivated by this, in Ref.~\cite{Cermeno:2021rtk} we studied the circular polarised signals of photons coming from DM interactions with cosmic ray electrons in the Galactic Centre (GC). We found that, although the circular polarisation asymmetry can reach up to $90\%$ at the distinctive peak present in the photon flux spectrum, the signal obtained does not seem to be detectable in the immediate future due to the low intensity of the photon flux. However, the high dependence of the photon flux and the circular polarisation asymmetry on the electron energy spectrum suggested that different sources, such as AGNs, could provide higher circular polarised fluxes and potentially lead to the detection of DM. Note that, although higher in general, the photon flux coming from the self-annihilation of DM will not produce a net circular polarisation asymmetry since the initial state for the interaction is a CP-eigenstate.

In this work, we explore the signals coming from leptophilic t-channel DM self-annihilation and scattering with electrons in AGNs. We calculate the circular polarisation asymmetry due to the interaction of our Majorana fermionic DM candidate with electrons of the jet radiating a photon in the final state. In particular, we consider a simplified model where the DM particle couples to right-handed electrons via a charged scalar mediator as it was done in Ref.~\cite{Cermeno:2021rtk}. 
The choice of this model is motivated by the fact that both DM self-annihilations and scatterings with electrons can provide monochromatic lines in the photon spectrum at the value of the DM mass and at the one of the mass splitting, respectively. This is the case because the leading order annihilation channel $\tilde{\chi} \tilde{\chi} \to e^+ e^-$ is velocity suppressed and therefore the dominant annihilation channels are the loop-induced $\gamma \gamma$ production and the virtual internal bremsstrahlung (VIB) $e^+ e^- \gamma$, characterised by line-like features in the small mass splitting regime  (see Refs.~\cite{Bringmann:2012vr, Garny:2013, Giacchino:2013bta, Okada:2014zja, Garny:2015wea, Kopp:2014, Cermeno:2021rtk}). Besides, the scattering will be a source of circularly polarised photons. 

In terms of the astrophysical source, we focus on Cen A.  This object is the closest active galaxy powered by its AGN, the estimated distance to the Earth is $d_{\text {AGN }} \sim 3.8$ Mpc (redshift $\sim 0.00183$) \cite{Abdalla:2018agf}, and the highest flux radio galaxy detected in hard X ray
and gamma ray bands \cite{Abdalla:2018agf}.

The paper is structured as follows. In Section~\ref{sec:flux}, we describe the theoretical framework for the calculations, focusing on the modeling of the DM density profile and the electron energy spectrum in the jet. The DM model used is also briefly described. In Section~\ref{sec:results}, we present the results of our study assuming two potential scenarios.
Firstly, we assume that the gamma ray photons measured by Fermi LAT and HESS coming from the core of Cen A are due to synchrotron self-Compton (SSC) radiation by two emitting zones, as it was modelled in Ref.~\cite{Abdalla:2018agf}. We show that, in this scenario, a huge range of our parameter space is excluded by the self-annihilation of our DM candidate, i.e $\tilde{\chi} \tilde{\chi} \rightarrow e^- e^+ \gamma$ and $\tilde{\chi} \tilde{\chi} \rightarrow \gamma \gamma$, if the DM density spike is present. These constraints are derived for the first time for leptophilic t-channel DM. We also calculate the flux of photons coming from the DM-electron scattering in the jet, $\tilde{\chi} e^- \rightarrow \tilde{\chi} e^- \gamma$, and its circular polarisation asymmetry.
From a completely different assumption, similarly to what it is done in Ref.~\cite{Brown:2016sbl}, modelling the background with a broken power law spectrum, we find an explanation for the excess of photons measured by Fermi-LAT between $2.5$ GeV and \C{$300$ GeV} with DM self-annihilation. For this analysis we consider both the prompt emission of photons coming from $\tilde{\chi}\tilde{\chi} \rightarrow e^- e^+ \gamma$, $\tilde{\chi}\tilde{\chi} \rightarrow \gamma \gamma$, $\tilde{\chi}\tilde{\chi} \rightarrow \gamma Z$ and the synchrotron radiation of the electrons produced by the 2 to 3 annihilation.
Finally in Section~\ref{sec:conclusion} we summarise our conclusions.

\section{Photon signatures from Cen A}
\label{sec:flux}

As it is illustrated in Ref.~\cite{Cermeno:2021rtk}, the flux of photons coming from a P violating interaction between DM and electrons can provide high circular polarisation asymmetries. This feature can be exploited to obtain information about the DM nature and its interactions with ordinary matter as well as to learn about the DM and electron distributions. However, the low intensity of the total flux of photons coming from these interactions in the GC makes these signals difficult to be observed. Here, we calculate the circular polarised flux of photons coming from DM particles of mass $m_{\tilde{\chi}}$ scattering off electrons in the AGN jet of Cen A, where both the density of DM and electrons are higher than in the GC. In particular, we consider the model studied in Ref.~\cite{Cermeno:2021rtk}, which is briefly described in Section II C, where the $\tilde{\chi} e^- \rightarrow \tilde{\chi} e^- \gamma_{\pm}$ scattering is resonant ($\pm$ indicates the positive or negative circular polarisation of the photon) and provides a peak in the photon spectrum for energies equal to the value of the mass splitting. This flux can be written as
\begin{equation}
\frac{d \Phi_{\gamma, \pm}}{d E_{\gamma}}=\frac{ \delta_{D M}}{m_\chi} \int d E_{e} \left(\frac{1}{d_{A G N}^{2}} \frac{d \phi_{e}^{A G N}}{d E_{e}}\right)  \frac{d^2\sigma_\pm}{d\Omega_\gamma dE_\gamma} (\theta_0,E_\gamma ),
\label{flux}
\end{equation}
where the last term is the differential cross section for the processes radiating photons with positive and negative circular polarisation. $\Omega_\gamma$ is the solid angle between the emitted photon and the incoming electron (with $\theta_\gamma$ the polar coordinate that is fixed at $\theta_0$ based on the position of the AGN with respect the line of sight), and $E_e$ and $E_\gamma$ are the incoming electron and the outgoing photon energies. 

In order to obtain the flux given by Eq.~\eqref{flux} we need to calculate three factors.
The first factor,
\begin{equation}
\delta_{D M} \equiv \int_{r_{m i n}}^{r_{0}} \rho_{D M}(r) d r,
\label{deltadm}
\end{equation}
is the integral of the DM density profile, $\rho_{D M}(r)$, over the distance from the center of the AGN, \C{along the direction of the jet}. The integration limits, $r_{m i n}$ and $r_{0}$,
are the minimum distance from the AGN center at which the scattering process we study takes place and the distance at which the AGN jet fades, respectively. Previous studies \cite{Gorchtein:2010xa} have found that, while results depend \C{sensitively} on the value of $r_{m i n}$, the actual value of $r_{0}$ plays little or no role (since DM density profiles typically fall off steeply with increasing radius).

The second factor involves the energy spectrum of the electron, $\frac{d \phi_{e}^{A G N}}{d E_{e}}$, that, as we have previously mentioned, has an important impact on the final results, and the AGN distance, $d_{\text {AGN }}$. 

Finally, the third factor depends upon the DM particle model and it involves the cross section at a scattering angle $\theta_{0}$ between the direction of the AGN jet and the line of sight. 

Apart from this signal, given the high DM density spike in the center of Cen A considered in this work, we expect a flux of photons coming from the annihilation of DM. 
The flux of such photons can be estimated as
\begin{equation}
\frac{d \Phi^{ann}}{d E_{\gamma}}=\frac{d N_{\gamma}}{d E_{\gamma}} \frac{\langle\sigma v\rangle}{8 \pi m_{\tilde{\chi}}^{2} d_{A G N}^{2}} \int_{r_{m i n}}^{r_{0}} d r 4 \pi r^{2} \rho_{D M}^{2}(r), 
\label{eq:ann}
\end{equation}
where $\langle\sigma v\rangle$ is the total annihilation cross section and $d N_{\gamma} / d E_{\gamma}$ is the differential photon
spectrum per annihilation event for the particular DM model under consideration. 

In order to compute this quantity we made use of the approximation 
\begin{equation}
\int_{\Delta \Omega} \int_{\rm l.o.s} \rho_{DM}^2(r) ds \, d\Omega  \approx \frac{4 \pi}{d_{AGN}^2}  \int_{4R_S}^{r_0} r^2 \rho_{DM}^2(r) dr \, ,
\end{equation}
where $s$ is the radial coordinate along the line of sight (l.o.s), $r=\sqrt{d_{AGN}^2+s^2-2d_{AGN} \, s \, \rm cos\; \rm \theta}$ and $\Delta \Omega$ the solid angle of observation. This approach is valid because $d_{AGN} \gg r_0$.

Note that, contrary to the ones coming from the scattering, the photons produced via DM annihilation will not be circularly polarised.

\subsection{DM density}

One of the potentially interesting features of some AGNs is the fact that, as depicted in Ref.~\cite{Gondolo:1999ef}, they can be characterised by the presence of a dense central spike of DM. In particular, this is expected to happen if DM particles are collisionless and the BH grows adiabatically by amassing stars, gas and DM. In this scenario, given an initial DM distribution such as $\rho (r) \propto \rho_0 \left(r/r_0\right)^{-\gamma}$, the functional form of the final DM density is given by
\begin{equation}
\rho_{\mathrm{DM}}(r)=\frac{\rho_{\mathrm{sp}}(r) \rho_{\mathrm{sat}}}{\rho_{\mathrm{sp}}(r)+\rho_{\mathrm{sat}}} \, ,   
\end{equation}
where the radial dependence of the profile is given by $\rho_{\mathrm{sp}}(r)$ and $\rho_{\mathrm{sat}}$ is the saturation density. The latter is the maximum density allowed by DM annihilation and reads
\begin{equation}
\rho_{\text {sat}} \simeq \frac{m_{\tilde{\chi}}}{\langle\sigma v\rangle  t_{B H}} \, ,
\end{equation}
where $t_{B H}$ is the age of the BH, $m_{\tilde{\chi}}$ is the DM mass and $\langle\sigma v\rangle$ the velocity averaged annihilation cross section. The radial dependence is instead given by
\begin{equation}
\rho_{\mathrm{sp}}(r)=\rho_R g_\gamma(r) \left(\frac{R_{\mathrm{sp}}}{r}\right)^{\gamma_{\mathrm{sp}}} \, ,   
\label{rho_radial}
\end{equation}
with
\begin{align*}
\rho_R & =\rho_{0}\left(\frac{R_{\mathrm{sp}}}{r_{0}}\right)^{-\gamma} \qquad  g_\gamma(r) \simeq \left(1-\frac{4 R_{S}}{r}\right)^{3} \\
\gamma_{s p} & =\frac{9-2 \gamma}{4-\gamma} \qquad \qquad \; R_{s p} =\alpha_{\gamma} r_{0}\left(\frac{M_{B H}}{\rho_{0} r_{0}^{3}}\right)^{1 /(3-\gamma)} \, .
\end{align*}
Specifically, $g_\gamma(r)$ accounts for the capture of DM in the BH, note that no stable orbit is allowed for non-relativistic particles below $4 R_S$~\cite{Sadeghian:2013laa, Kavanagh:2020cfn}, being $ R_S$ is the Schwarzschild radius. Regarding the rest of factors, $\gamma_{s p}$ is the slope of the density profile in the spike and $R_{s p}$ is the radius of the spike, which depends on the BH mass $M_{BH}$ and the normalization constant $\alpha_\gamma$.
\begin{figure*}[t!]
	\centering
	\includegraphics[width=0.45\linewidth]{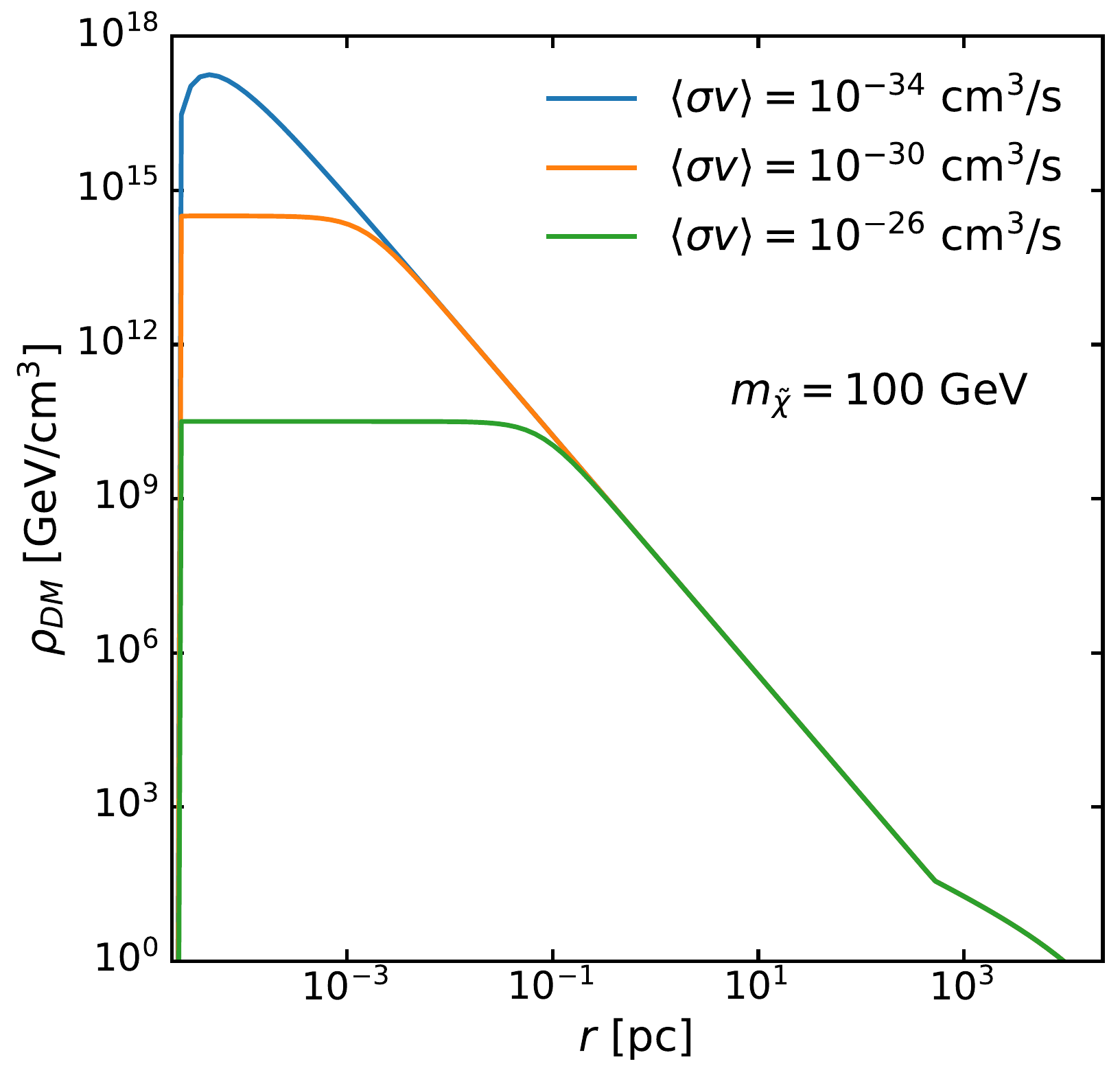}
	\includegraphics[width=0.47\linewidth]{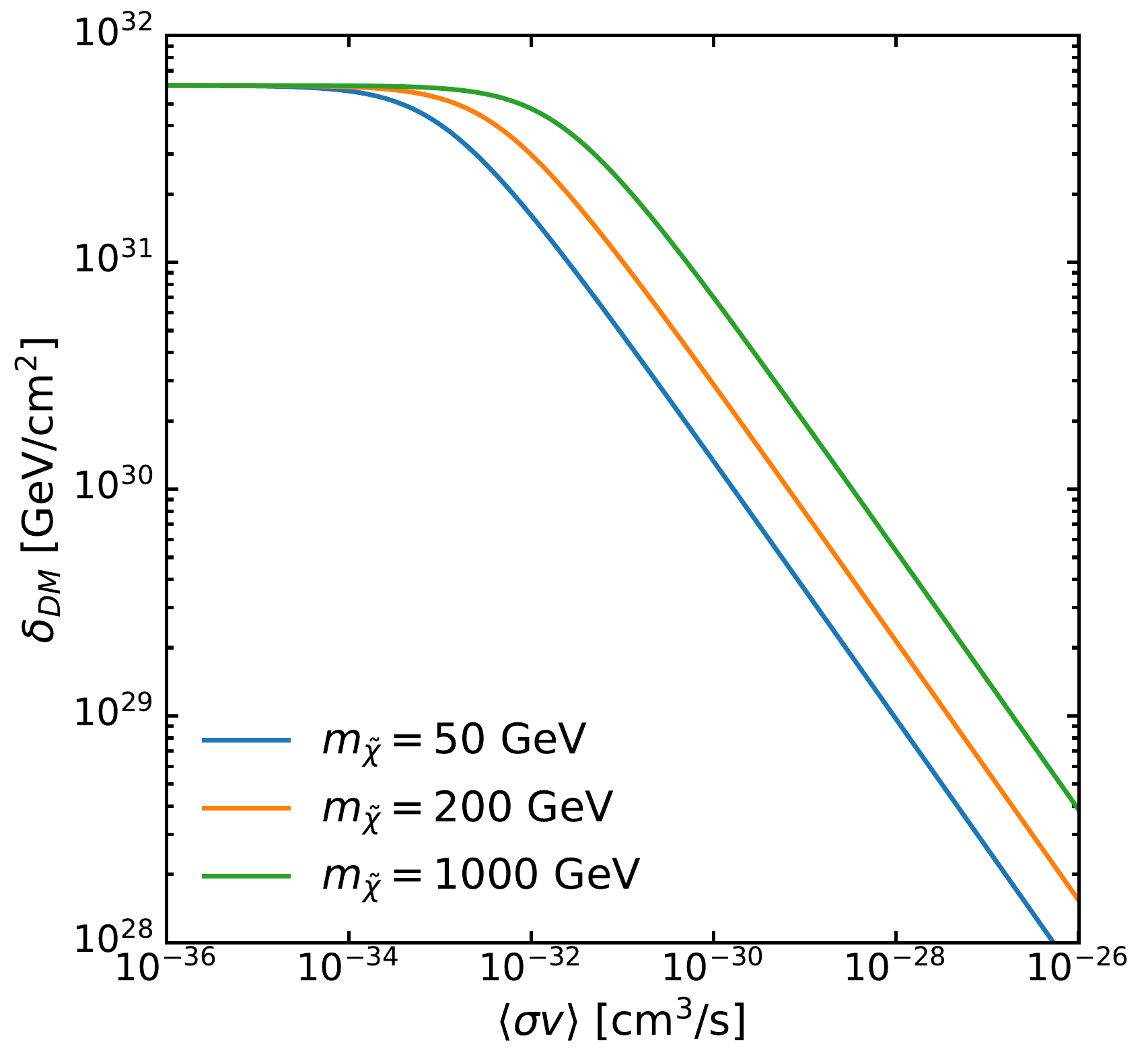}
	\caption{Left: DM density profile as a function of the radial distance to the AGN center in the scenario of a spike formed in the vicinity of the central BH of Cen A for different values of the velocity-averaged cross section $\langle\sigma v\rangle$ and fixing $m_{\tilde{\chi}}=100$ GeV. Right: The $\delta_{DM}$ quantity that enters the flux calculation as a function of $\langle\sigma v\rangle$ for different DM mass values.}
	\label{fig:DMspike}
\end{figure*}

It is worth pointing out that the spike structure is expected to smooth down because of dynamical relaxation caused by the scattering of DM with stars, leading to a power law DM profile $\rho_{DM}(r) \sim r^{-\frac{3}{2}}$~\cite{Gnedin:2003rj}. However, this is supposedly not the case for Cen A~\cite{Lacroix:2015lxa, Lacroix:2016qpq, Brown:2016sbl}, being a dynamically young AGN: its relaxation time is expected to be $\sim 10^2$ Gyr, which is larger than the Hubble
time \C{$\sim 14 \; \rm Gyr$} and therefore dynamical heating by stars is inefficient. As a result, a spike formed in early times could have survived until now. \C{Nevertheless, it is worth stressing that no direct evidence of spikes has been found so far and their existence is debatable since other aspects can disrupt the spike. For example, if the growth of the BH was instantaneous, the adiabatic treatment would not be suitable and the DM density profile would scale instead as $\rho_{DM}(r) \propto r^{-\frac{4}{3}}$~\cite{Ullio:2001fb}. Mergers between halos containing supermassive BHs (SMBHs) can also destroy DM spikes by expelling DM particles from the center due to the kinetic heating induced by the SMBH binaries created. In this case, the DM density profile would behave as $\rho_{DM}(r) \propto r^{-\frac{1}{2}}$~\cite{Merritt:2002vj}. The same profile is expected if the BH did not grow exactly at the center of the DM halo, but may have grown instead from a BH seed brought in by a merger of progenitor halos, and then spiraled
	in to the center~\cite{Ullio:2001fb}. In this work, apart for the adiabatic growth of the SMBH, we assume that this has occurred in the center of the DM halo and that the halo did not undergo a merger. } 

For a radius bigger than $R_{sp}$, the DM density is described by a standard power law behaviour, which we describe with a Navarro Frenk and White (NFW) profile, such as
\begin{equation}
\rho_{\text {halo }}(r) = \rho_0 \left(\frac{r}{r_0}\right)^{-\gamma}\left(1+\frac{r}{r_0}\right)^{-2} \, ,
\end{equation}
where $\gamma = 1$ and, therefore, $\gamma_{sp} = 7/3$.

Given all of the above, the functional form of the density considered reads
\begin{equation}
\rho_{DM}(r)=\left\{\begin{array}{ll}
0 & r<4 R_{\mathrm{S}} \\
\frac{\rho_{\mathrm{sp}}(r) \rho_{\mathrm{sat}}}{\rho_{\mathrm{sp}}(r)+\rho_{\mathrm{sat}}} & 4 R_{\mathrm{S}} \leq r<R_{\mathrm{sp}} \; . \\
\rho_{\text {halo }}(r) & r \geq R_{\mathrm{sp}}
\end{array}\right. 
\end{equation}

In order to fix the parameters $\rho_0$ and $R_{sp}$ we require that the DM density satisfies the following equations
\begin{align}
\int_{4 R_{S}}^{10^{5} R_{S}} 4 \pi r^{2} \rho_{D M}(r) d r &\lesssim \Delta M_{B H}\, , \label{eq:BHmass1}\\
\int_{4 R_{S}}^{50 \;\rm kpc} 4 \pi r^{2} \rho_{D M}(r) d r &\lesssim 10^{12} M_\odot \, .
\label{eq:BHmass2}
\end{align}
The first equation is requiring that the DM mass in the vicinity of the BH is not bigger than the uncertainty on the BH mass itself, $\Delta M_{B H}$, and the second equation is requiring that the total DM mass in the galaxy does not exceed the mass of the galaxy, $\sim 10^{12} M_\odot$. In particular, for Cen A, following Refs.~\cite{Lacroix:2016sjj,Neumayer_2010}, we have 
\begin{align*}
t_{\rm BH}&=10^{10} \, \mathrm{yr} \qquad \qquad \Delta M_{\mathrm{BH}}=3 \times 10^7 M_{\odot} \\ M_{BH}&=5.5 \times 10^7 M_{\odot} \qquad \quad
R_S = 5 \times 10^{-6} \, \mathrm{pc}\\ \alpha_\gamma&=0.1 \qquad \qquad \qquad \quad \; \; \; r_0=20 \, \mathrm{kpc} \, .
\end{align*}
Plugging these parameters in the above equations we find $\rho_0 \sim 1 \; \rm GeV/cm^3$ and $R_{\rm sp} =10^8 R_S$.

On the left side of Fig.~\ref{fig:DMspike}, we show the DM density profile as a function of the radial distance to the center of the AGN for different values of the velocity-averaged annihilation cross section and a DM candidate of mass $m_{\tilde{\chi}}=100$~GeV. As it can be seen from the plot, the smaller the cross section is, the bigger the spike in the vicinity of the BH is. On the right side of Fig.~\ref{fig:DMspike}, we instead show the values of $\delta_{DM}$, defined in Eq.~\eqref{deltadm}, for different DM masses as a function of $\langle\sigma v\rangle$. It is interesting to note that, while for higher values of the velocity-averaged annihilation cross section there are differences in $\delta_{DM}$, below $\left<\sigma v\right> \sim 10^{-34}\; \rm cm^3/s$ the quantity plateaus around $6\cdot10^{31}$ GeV/cm$^2$ independently of the DM mass.

In the event that no dense spike of DM has been formed in Cen A, we consider a scenario in which the DM density profile is given by the classic NFW functional form. In particular, we take
\begin{equation}
\rho_{DM}(r)=\left\{\begin{array}{ll}
0 & r<4 R_{\mathrm{S}} \\
\rho_{\rm sat} & 4 R_{\mathrm{S}} \leq r<R_{\rm sat} \\
\rho_{\text {halo }}(r) & r \geq R_{\rm sat}
\end{array}\right. 
\end{equation}
with $R_{\rm sat}$ being the radius below which the DM density saturates, which is given by the formula \cite{Lacroix:2016sjj}
\begin{equation}
R_{\rm sat} = r_0 \left(\frac{\rho_0 \langle\sigma v\rangle t_{BH}}{m_{\tilde{\chi}}}\right)^{\frac{1}{\gamma}} \, .
\end{equation}

Also for this case, we verify that the Eqs.~\eqref{eq:BHmass1} and~\eqref{eq:BHmass2} are fulfilled. If DM exhibits the latter density profile, we find $\delta_{DM}\sim 10^{24}$ GeV/cm$^2$, which is $7-8$ orders of magnitude smaller than the one obtained with the spike profile. Therefore, considerably lower photon fluxes are expected if no spike is present.

\begin{center}
	\begin{table*}[t!]
		\begin{tabular}{|c|c|c|}
			\hline 
			\text{Parameter } & \text{the 1st SSC zone} & \text{the 2nd SSC zone} \\
			\hline 
			$\delta$ & 1.0 & 1.0  \\
			$\theta_0$ & $30^{\circ}$ & $30^{\circ}$ \\
			B (G) & 6.2 & 17.0 \\
			$R_{\mathrm{b}}$ (cm)& $3.0 \times 10^{15}$ & $8.8 \times 10^{13}$\\
			\hline 
			$s_{1}$ & 1.8 & 1.5\\
			$s_{2}$ & 4.3 & 2.5\\
			$\gamma'_{\min }$ & $3 \times 10^{2}$ & $1.5 \times 10^{3}$\\
			$\gamma'_{\max}$ & $1 \times 10^{7}$ & $1 \times 10^{7}$ \\
			$\gamma'_{\text {brk }}$ & $8.0 \times 10^{2}$ & 3.2 $\times 10^{4}$ \\
			$L_e$ $\left(\mathrm{erg} \; \mathrm{s}^{-1}\right)$ & $3.1\times 10^{43}$ & $3\times 10^{40}$ \\
			\hline
		\end{tabular}
		\caption{Parameters related to the electron energy spectrum for Cen A \cite{Abdalla:2018agf}}
		\label{table1}
	\end{table*}
\end{center}
\subsection{The electron energy spectrum in the jet}
\label{subsec:jet}
In this section we discuss the modeling of the electron energy spectrum in the AGN jet, \ie the second factor in the photon flux of Eq.~\eqref{flux}.
As shown in Refs.~\cite{Gorchtein:2010xa, Huang:2011dg, Gomez:2013qra}, the geometry details of the AGN jet are not very important for our calculation. However, modeling the energy spectrum of electrons in the jet accurately is crucial. Here we consider the blob geometry used in Refs.~\cite{Gorchtein:2010xa, Huang:2011dg, Gomez:2013qra}, where electrons move isotropically in the blob frame with a power law energy distribution. Furthermore, the blob moves with respect to the central BH with a bulk Lorentz factor towards the observer in a jet given by $\Gamma_B = (1-\beta_B^2 )^{-\frac{1}{2}}$  with an angle $\theta_0$ to the line of sight. Therefore, the emission is Doppler-shifted with a Doppler factor $\delta=\left[\Gamma_B(1-\beta_B \; \textrm{cos}\;\theta_0)\right]^{-1}$. In particular, for Centaurus A, $\Gamma_B \sim 7$  and $\delta \sim 1$ \cite{Abdalla:2018agf}. Based on the observations of emitted gamma rays from the core of Cen A by the Fermi LAT and HESS experiments, the distribution for the relativistic particles in the jet (in the blob frame) can be described as a broken power law~\cite{Abdalla:2018agf}
\begin{eqnarray}
& &\frac{d \phi_{e}^{\mathrm{AGN}}}{d \gamma^{\prime}}\left(\gamma^{\prime}\right)=\frac{1}{2} k_{e} \gamma^{\prime-s_{1}}\left[1+\left(\frac{\gamma^{\prime}}{\gamma_{b r}^{\prime}}\right)^{\left(s_{2}-s_{1}\right)}\right]^{-1} \\ \nonumber & & \textrm{for} \quad  \gamma_{\min }^{\prime}<\gamma^{\prime}<\gamma_{\max }^{\prime}
\end{eqnarray}
with $\gamma '=\frac{E'_e}{m_e}$, being $E'_e$ the electron energy in the blob frame and $m_e$ its mass. See Ref.~\cite{Finke:2008pe} and the Appendix A from Ref.~\cite{Lacroix:2016sjj} for more details. The parameters $s_{1}, s_{2}, \gamma_{b r}^{\prime}, \gamma_{\min }^{\prime}$ and $\gamma_{\max }^{\prime}$ are taken from the model used in Ref.~\cite{Abdalla:2018agf}, which is described by two SSC emission zones, since a single zone SSC model is not able to adequately account for the overall core spectral energy distribution (SED) of Cen A. The values used for these parameters are reported in Table 1. 
Note, however, that the gamma ray data used in Ref.~\cite{Abdalla:2018agf} should be treated as upper limits since part of this emission for $E_\gamma \gtrsim 100 \; \rm GeV $ arises on large scales, as it has been pointed out in Ref.~\cite{Hess:2020tvd}.

The normalization constant $k_{e}$ can be determined from the jet power in electrons, which is defined in the BH frame as
\begin{equation}
L_{e}=\int_{-1}^{1} \frac{d \mu}{\Gamma_{B}\left(1-\beta_{B} \mu\right)} \int_{\gamma_{\min }}^{\gamma_{\max }} d \gamma \, m_{e} \gamma \frac{d \phi_{e}^{A G N}}{d \gamma}\left(\gamma, \mu\right),   
\end{equation}
with $\mu=\cos \theta$ and $\gamma=E_e / m_{e}$, being $E_e$ the energy of the electron. Similarly, in the blob frame $\mu^{\prime}=\cos \theta^{\prime}$. Note that the jet power in electrons can be written as $L_e= \pi R_b^2 \beta_B u_e$, where $u_e$ is the energy density for a cylindrical jet region of radius $R_b$. The quantities in the two frames are related by the blob velocity $\beta_{B}$ and boost $\Gamma_{B}$ as
\begin{equation}
\mu^{\prime}=\frac{\mu-\beta_{B}}{1-\beta_{B} \mu}, \quad \gamma^{\prime}=\gamma \Gamma_{B}\left(1-\beta_{B} \mu\right),    
\end{equation}
or equivalently,
\begin{equation}
\mu=\frac{\mu^{\prime}+\beta_{B}}{1+\beta_{B} \mu^{\prime}}, \quad \gamma=\frac{\gamma^{\prime}}{\Gamma_{B}\left(1-\beta_{B} \mu\right)}.    
\end{equation}
In this way
\begin{eqnarray}
& & \int_{-1}^{1} d \mu^{\prime} \int_{\gamma_{\min }^{\prime}}^{\gamma_{\max }^{\prime}} d \gamma^{\prime} \frac{d \phi_{e}^{\mathrm{AGN}}}{d \gamma^{\prime}}\left(\gamma^{\prime}\right)=\\ \nonumber & & \int_{-1}^{1} \frac{d \mu}{\Gamma_{B}\left(1-\beta_{B} \mu\right)} \int_{\gamma_{\min }}^{\gamma_{\max }} d \gamma \frac{d \phi_{e}^{\mathrm{AGN}}}{d \gamma}\left(\gamma, \mu\right),   
\end{eqnarray}
where the limits on the integral over $\gamma$ are given by
\begin{equation}
\gamma_{\min }=\frac{\gamma_{\min }^{\prime}}{\Gamma_{B}\left(1-\beta_{B} \mu\right)}, \quad \gamma_{\max }=\frac{\gamma_{\max }^{\prime}}{\Gamma_{B}\left(1-\beta_{B} \mu\right)}.    
\end{equation}
Therefore, the distribution for the relativistic electrons in the jet in the BH frame can be written as
\begin{equation}
\frac{d \phi_{e}^{\mathrm{AGN}}}{d \gamma}\left(\gamma, \mu\right)=\frac{1}{2} \frac{k_{e}\left[\gamma \Gamma_{B}\left(1-\beta_{B} \mu\right)\right]^{-s_{1}}}{1+\left(\gamma\left(\Gamma_{B}\left(1-\beta_{B} \mu\right)\right) / \gamma_{b r}^{\prime}\right)^{s_{2}-s_{1}}} .   
\end{equation}
Using this, the electron energy spectrum which appears together with the distance to the AGN $d_{\rm AGN}$ in the second factor that we need to compute the flux of Eq.~\eqref{flux} reads
\begin{equation}
\frac{d \phi_{e}^{\mathrm{AGN}}}{d E_{e}}= \frac{1}{ m_e} \int_{\mu_0}^{1} \frac{d \mu}{\Gamma_{B}\left(1-\beta_{B} \mu\right)} \frac{d \phi_{e}^{\mathrm{AGN}}}{d \gamma}\left(\gamma, \mu\right),
\label{eq:eespectrum}
\end{equation}
where $\mu_0$ parameterizes the jet collimation. In the following, we adopt a value of $\mu_0=0.9$ to ensure
a highly-collimated jet.

\subsection{The DM model}
\label{subsec:model}
In order to obtain quantitative results, we consider a specific leptophilic DM simplified model, following the works in Refs.~\cite{Profumo:2011jt, Kopp:2014, Garny:2015wea, Cermeno:2021rtk}. The Lagrangian of the model of interest is given by
\begin{widetext}
	\begin{equation}
	\mathcal{L}_{DM} = i \bar{\psi}_{\tilde{\chi}}(\slashed{D} - m_{\tilde{\chi}})\psi_{\tilde{\chi}} + D_\mu \varphi^\dagger D^\mu \varphi - m_\varphi \varphi^\dagger \varphi + ( a_R \, \bar{e}_R \, \psi_{\tilde{\chi}} \, \varphi + h.c.) \, ,  
	\end{equation}
\end{widetext}
where the SM has been extended with two additional degrees of freedom: a Majorana DM candidate $\tilde{\chi}$ and a scalar mediator $\varphi$ which couples to right-handed electrons via the coupling $a_R$. \C{Since DM is a singlet of the SM gauge group}, the scalar mediator has to carry all the charges of the SM right-handed electron and is therefore charged. The parameter space of the model is three dimensional and completely specified by the parameters $\{m_{\tilde{\chi}}, m_\varphi, a_R \}$. It is worth noting that in order to ensure the stability of the DM candidate, its mass cannot exceed the one of the mediator particle. Consequently, the further constrain $m_\varphi > m_{\tilde{\chi}}$ is imposed on the parameter space.

The choice of considering a t-channel model is related to the fact that these kind of models
exhibit a resonant enhancement of the DM-electron interactions, which can be exploited to probe the small mass splitting regime $\Delta M = m_\varphi - m_{\tilde{\chi}}\sim [0.1, 10]$ GeV and open a new window on the exploration of a region of the parameter space which is of difficult study in other experimental methods. Besides, we choose a parity violating interaction where DM particles only couple to right handed electrons in order to maximise the circular polarisation asymmetry. Additionally, being the model characterised by a p-wave suppressed annihilation into electrons, the distinctive annihilation photon signals are all line-like, coming from VIB and loop-induced $\gamma \gamma$ and $Z \gamma$.

We obtain our results through the use of a model implemented in FeynRules~\cite{Alloul:2013bka}, which is then used to compute analytical and numerical results with tools like FeynArts~\cite{Hahn:2000kx}, FeynCalc~\cite{feyncalc1, feyncalc2, feyncalc3},  \mgFull~\cite{Alwall:2014hca}, MadDM~\cite{Ambrogi:2018jqj, Arina:2020kko} and MicrOMEGAs~\cite{Belanger:2018ccd}.

For a detailed discussion on the current experimental limits on the parameter space of the model, we refer the reader to Ref.~\cite{Cermeno:2021rtk}.

\section{Results}
\label{sec:results}
In the following we present the results of our study. In particular, we consider two separate scenarios according to different assumptions on the measured flux of gamma rays from Cen A. In the first part of the analysis, we assume that the photon SED can be explained completely by astrophysical known sources, \ie two SSC jet-related components of the Cen A AGN, as it is done in Ref.~\cite{Abdalla:2018agf}, and therefore the detected photons can be considered as background for potential DM signals. In the latter scenario, we assume that while part of the SED is explained by SM processes in the AGN jets, DM is responsible for what looks like an excess of photons in the $[10-100]$ GeV energy window.

\subsection{Exclusion limits from DM self-annihilation and circularly polarised photon flux}

Following the assumptions for the SED done in Ref.~\cite{Abdalla:2018agf}, which we have used in Section~\ref{subsec:jet} to derive the energy distribution of relativistic electrons, we can now project constraints on the model that we are considering. In particular, we consider the photon flux from Ref.~\cite{Abdalla:2018agf} (Fig. 3) as our background. We can parametrise the flux of a monochromatic signal as
\begin{equation}
\frac{d\Phi^{ann}}{dE_\gamma} = \frac{\langle\sigma v\rangle }{2 m_{\tilde{\chi}}^{2} d_{A G N}^{2}} I_{\tilde{\chi}}(r_0) \delta(E_\gamma - m_{\tilde{\chi}}) \, \, ,
\end{equation}
where
\begin{equation}
I_{\tilde{\chi}}(r_0) = \int_{r_{m i n}}^{r_{0}} d r \, r^{2} \rho_{D M}^{2}(r) \, .
\end{equation}
Given that our DM model is characterised by an helicity suppression for the leading order $2\to2$ scattering, the dominant annihilation signal is given by the 3-body annihilation $\tilde{\chi} \tilde{\chi} \to e^+ e^- \gamma$ and the loop-induced process $\tilde{\chi} \tilde{\chi} \to \gamma \gamma$, which are both characterised by a line-like signal in the small mass splitting regime. In particular, in the non degenerate region of the parameter space, the annihilation into two photons is the dominating contribution, while for small mass splitting it is the VIB channel that dominates.
\begin{figure}[t!]
	\centering
	\includegraphics[width=0.9\linewidth]{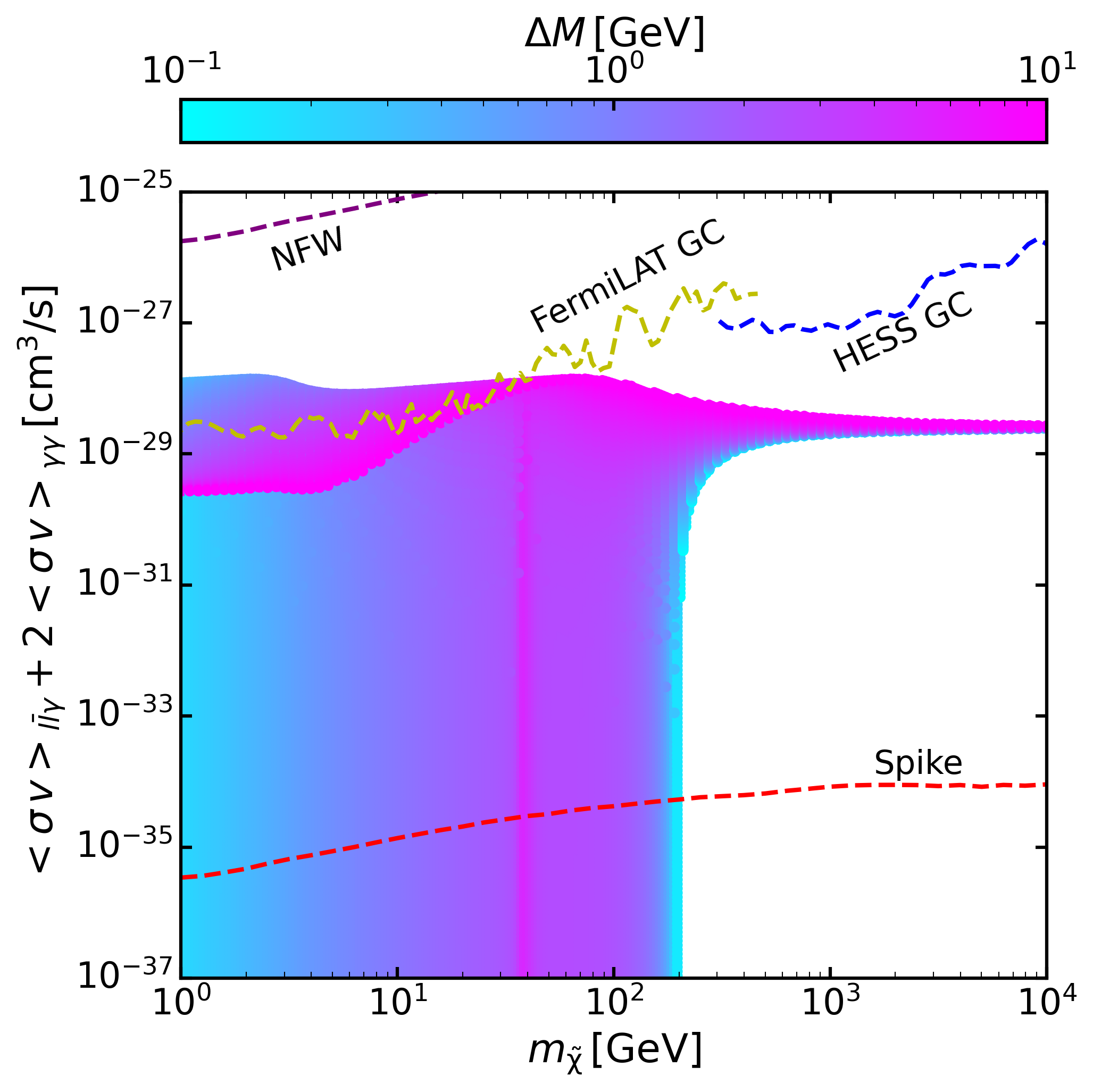}
	\caption{Exclusion limits on the total annihilation cross section as a function of the DM mass for different values of the mass splitting $\Delta M$ using the data of Fermi LAT and HESS. The red line is derived in the hypothesis of a DM spike being present in the center of Cen A while the purple one assumes that the DM density is described by the NFW profile. The data to derive this lines are taken from Ref.~\cite{Abdalla:2018agf}.
		The parameter space points showed in the plot are the ones that satisfy the DM relic density constraint. For comparison the constraints coming from the annihilation of DM in the GC for a NFW profile (taken from Refs.~\cite{Fermi-LAT:2015kyq, HESS:2018cbt}) are included (yellow and blue lines).}
	\label{fig:exclusions}
\end{figure}

In Fig.~\ref{fig:exclusions} we report the exclusion limits on $\langle\sigma v\rangle$ that we find, both in the hypothesis that a spike is present in the vicinity of the BH (red line) and in the hypothesis that the DM profile is described by a NFW distribution (purple line). The points shown in the plot are the ones that provide the correct relic density $\Omega_{\chi}=0.12$ (see Ref.~\cite{Cermeno:2021rtk} for details). In particular, we exclude points which would provide a photon flux higher than the one measured by Fermi-LAT and HESS, \ie the data showed in Fig. 3 of Ref.~\cite{Abdalla:2018agf}. Apart from the constraints coming from Cen A, we also plot for comparison the bounds on the model coming from the observation of GC photons by Fermi-LAT (yellow line) \cite{Fermi-LAT:2015kyq} and HESS (blue line) \cite{HESS:2018cbt}.

As it can be observed in Fig.~\ref{fig:exclusions}, while from the GC constraints are very mild due to the fact that the spike does not survive to date, in the presence of a DM spike in the inner region of Cen A, the vast majority of the parameter space would be excluded. In particular, no DM candidate with a mass higher that $200$ GeV survives. However, there is an infinite set of solutions at smaller masses which is characterised by co-annihilation driven freeze-out and evades the constraints for annihilation cross sections below $\langle\sigma v\rangle \sim 10^{-34}-10^{-35}$ cm$^3$/s.

On the other hand, in the hypothesis that the spike has not been formed, no constraint is imposed on the model. However, we find that this scenario is of lower interest given that no measurable signal can be detected in the near future, neither from annihilation nor from the scattering, and Cen A would not be competitive with observations from other astrophysical sources such as the GC. For this reason, we decide to not comment further on this scenario.

\begin{figure}[t!]
	\centering
	\includegraphics[width=0.98\linewidth]{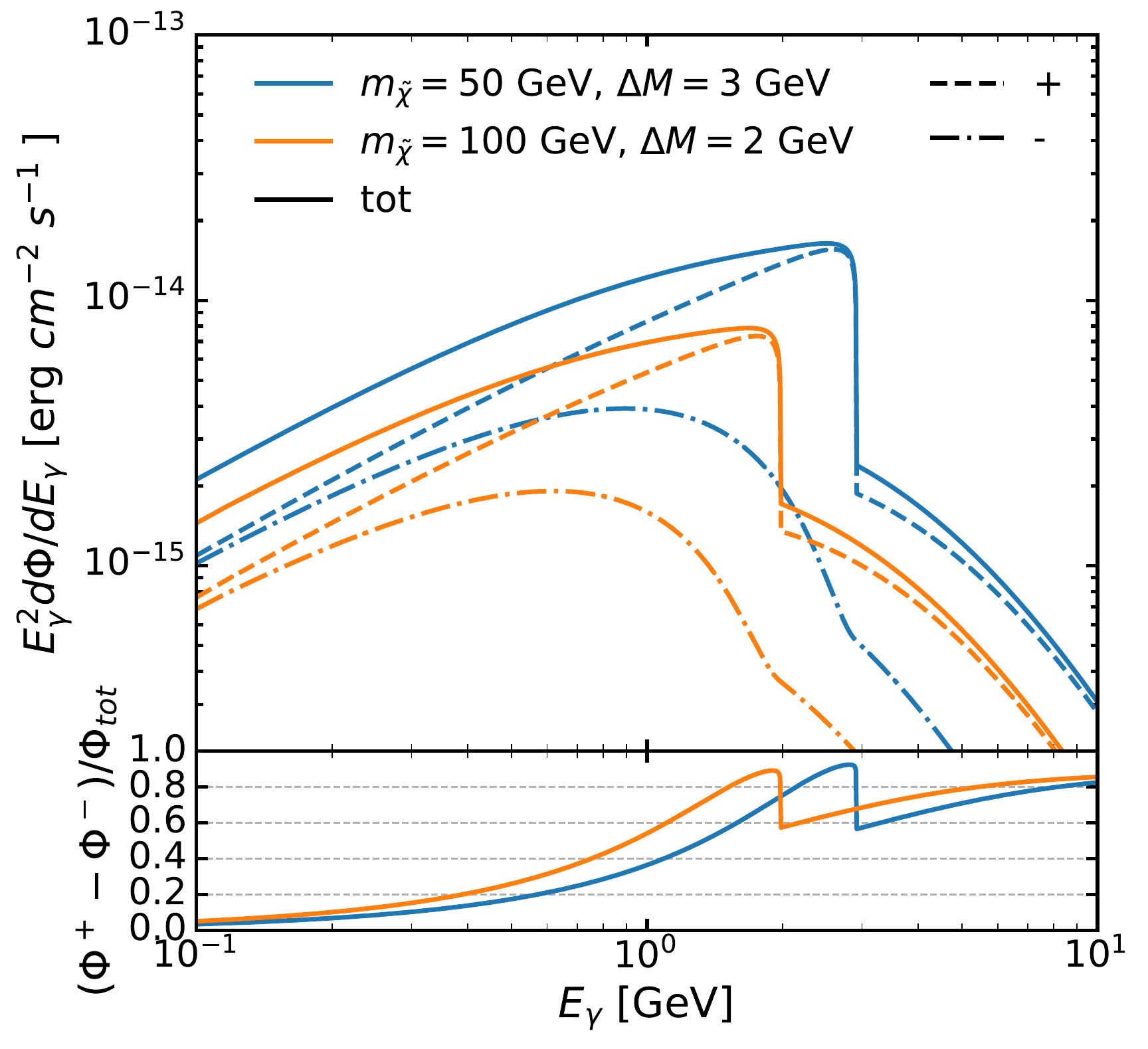}
	\caption{The predicted photon fluxes coming from the scattering of DM particles with masses of $50$ GeV (blue) and $100$ GeV (orange) with the electrons of the AGN jet of Cen A in the presence of a spike in the vicinity of the BH are shown. Dashed and dashed dotted lines correspond to the flux of photons with positive and negative circular polarisation respectively, and  solid lines to the sum of both contributions. In the lower panel, the circular polarisation asymmetry is reported}
	\label{fig:scattering}
\end{figure}

We now move on to discuss the possible detection of a circularly polarised signal from Cen A coming from the scattering of DM particles with electrons in the jet. From Fig.~\ref{fig:exclusions}, we infer that allowed DM candidates have masses that span from $40$ to $200$ GeV and a coupling $a_R \lesssim 10^{-2}$. Candidates with masses lower than $40$ GeV are not considered since $Z$-width constraints substantially exclude these scenarios, given that a new decay channel into a pair of mediator particles would open up. Interestingly, the mass splittings for the survived parameter space points are fixed by the relic density constraint and span from $\Delta M = 3 \, (1)$ GeV for a DM candidate with mass $m_{\tilde{\chi}}=50 \, (200)$ GeV. 

In Fig.~\ref{fig:scattering} we report the predicted fluxes for some benchmark allowed parameter space points, \C{i.e. $m_\chi= 50$ GeV, $\Delta M =3$ GeV in blue and $m_\chi= 100$ GeV, $\Delta M =2$ GeV in orange for $a_R \sim 10^{-2}$}. In dashed and dashed-dotted the flux of photons with positive and negative circular polarisation respectively can be found, while solid lines denote the total flux of photons, \ie,
\begin{equation}
\frac{d\Phi_{\gamma, tot}}{dE_\gamma}=\frac{d\Phi_{\gamma, +}}{dE_\gamma}+\frac{d\Phi_{\gamma, -}}{dE_\gamma}.    
\end{equation}
The calculation of the fluxes is performed in the narrow width approximation (NWA) and further details on the validity of it and the various kinematic features of the process are discussed in Ref.~\cite{Cermeno:2021rtk}.

From direct inspection of the plot, we observe that the expected flux of photons reaches values of $\sim 10^{-14}$ erg cm$^{-2}$s$^{-1}$ in the peak around $\Delta M \sim $ few GeVs. The intensity of this flux is around $2$ orders of magnitude smaller than the background and a detection in absence of a considerable advancement in the signal to background discrimination is unlikely in the near future.

It is worth noting though that, in order to obtain our results, we are using 
the parameters of Table~\ref{table1}, which completely determine the electron energy spectrum given by Eq.~\eqref{eq:eespectrum}. This means that we are fixing the jet power in electrons to $L_e = 3.1 \; 10^{43} \; \rm erg/cm^3$ for the 1st SCC zone and to $L_e = 3 \; 10^{40} \; \rm erg/cm^3$ for the 2nd SCC (red and blue lines of Fig. 3 of Ref.~\cite{Abdalla:2018agf}). On the contrary, the authors of Refs.~\cite{Gorchtein:2010xa, Huang:2011dg, Gomez:2013qra} use the fit for the electron energy spectrum given by the brown line of Fig. 5 of Ref.~\cite{Abdo_2010} but they consider a much higher value for the jet power in electrons, \ie the Eddington limit $L_e \sim 10^{45} \; \rm erg/cm^3$, instead of the one needed to fit the data with the model they used, which is $L_e = 7 \; 10^{40} \; \rm erg/cm^3$. Since the value of $L_e$ fixes the normalization constant $k_e$, assuming a jet power in electrons of the order of the Eddington limit would enhance our expected photon flux coming from DM-electron scatterings by two orders of magnitude, meaning that the flux would be of the order of the sensitivity of the experiment or even higher. However, an increase in the jet power has to be followed by a consistent reparametrisation of the jet, otherwise the flux of photons coming from the SSC of electrons would exceed the measured photon flux.

Regarding the circular polarisation asymmetry, defined by
\begin{equation}
\frac{\Phi^+-\Phi^-}{\Phi_{tot}} \equiv \frac{\frac{d \Phi_{\gamma, +}}{d E_{\gamma}}-\frac{d \Phi_{\gamma, -}}{d E_{\gamma}}}{ \frac{d \Phi_{\gamma, tot}}{d E_{\gamma}}},   
\end{equation}
looking at the lower panel of Fig.~\ref{fig:scattering}, we notice that the degree of asymmetry at the peak is almost $100\%$. Since astrophysical sources could not mimic this high circular polarisation asymmetries, a measurement of this will provide a proof on new physics. However, the only efficient methods developed to date for circular polarization measurements are based on Compton scattering. These techniques, which exploit the correlation of the outgoing electron spin with the initial photon helicity, measure the secondary asymmetries caused by a primary gamma ray flux. This means that in practise the sensitivity with respect to the one needed to measure the total flux decreases by a factor $A \sqrt{\epsilon}_p$, where $A \sim 10\%$ is the typical asymmetry in the secondary particle spectra expected from a $100\%$ polarised gamma ray flux and $\epsilon_p$ the efficiency of the detector for useful events. See Refs.~\cite{Cermeno:2021rtk, Elagin:2017cgu} for more details on this. Therefore, this suggest that novel techniques should be developed in order to exploit the polarisation of the signal with the objective of disentangling the background.
\begin{figure}[t!]
	\centering
	\includegraphics[width=0.98\linewidth]{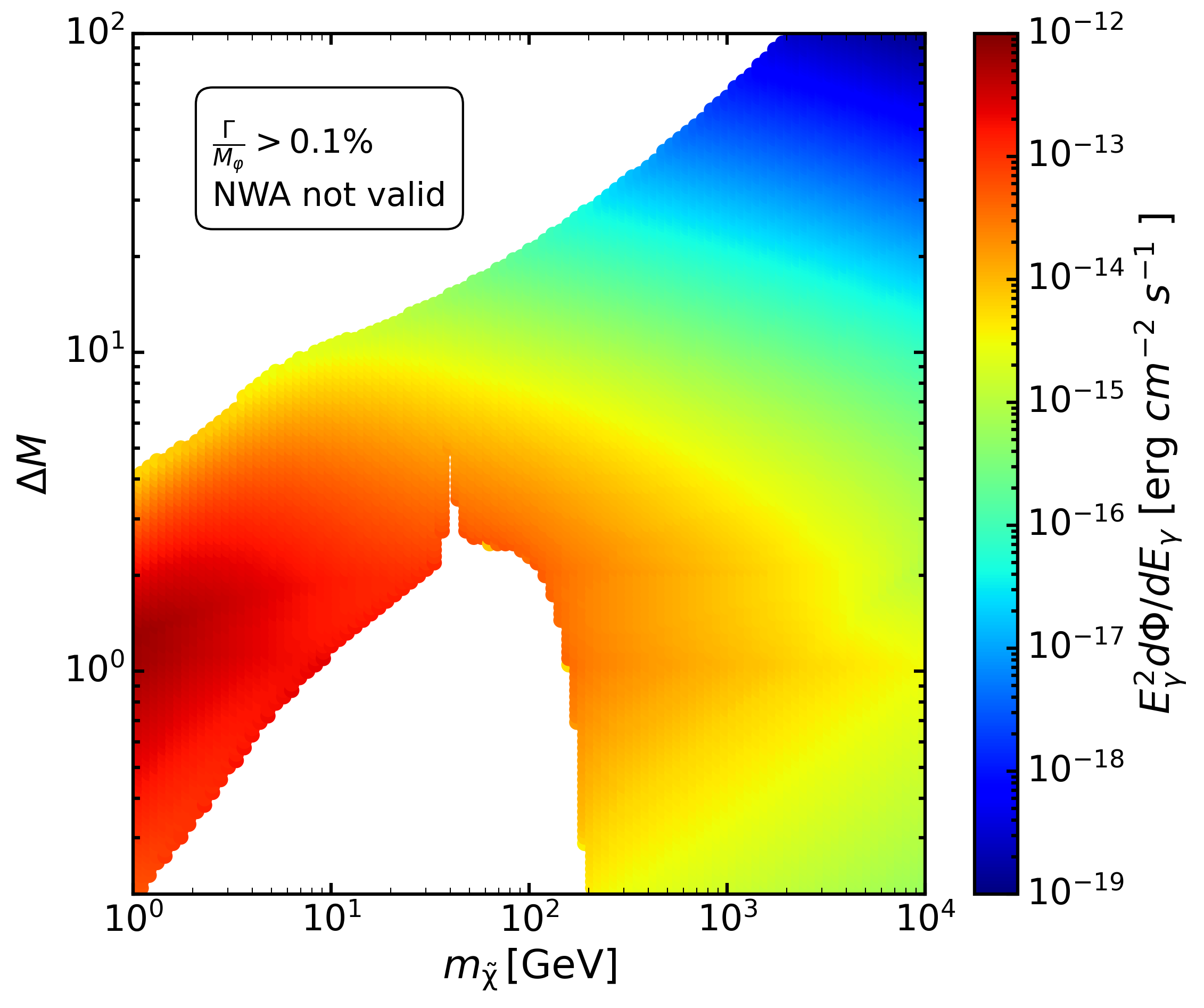}
	\caption{Heatmap of the peak value of the fluxes coming from scattering of DM with the electrons of the AGN jet in the scenario of a DM density in the core. The parameter points displayed are the ones that provide the correct relic abundance and for which the NWA is valid.}
	\label{fig:heatmap}
\end{figure}

For completeness, in Fig.~\ref{fig:heatmap} we show a heatmap highlighting the peak value of the flux from the scattering with electrons for all the parameter space points which provide the correct relic density abundance. Points for which the NWA is not valid are not shown. Notice that the position of the peak of the flux is at $\Delta M$. We observe that the maximum values are obtained for low mass DM candidates, which are however unfortunately ruled out by $Z$ boson decay constraints. \C{In particular, for the parameter space points allowed by the $Z$ boson decay constraint, the flux in the peak cannot exceed $10^{-13}\; \rm erg\; cm^{-2}\;s^{-1}$, being therefore lower than the sensitivity of Fermi-LAT to measure this signal in the upcoming years ($\sim 5 \; 10^{-13}\; \rm erg\; cm^{-2}\; s^{-1}$ in 30 years of data collection). This means that, in order to detect this signal, we would need a coupling 3 times bigger than the one providing the correct relic abundance for our model.}

\subsection{Fit of the photon excess with DM self-annihilations}
\begin{figure*}[t!]
	\centering
	\includegraphics[width=0.46\linewidth]{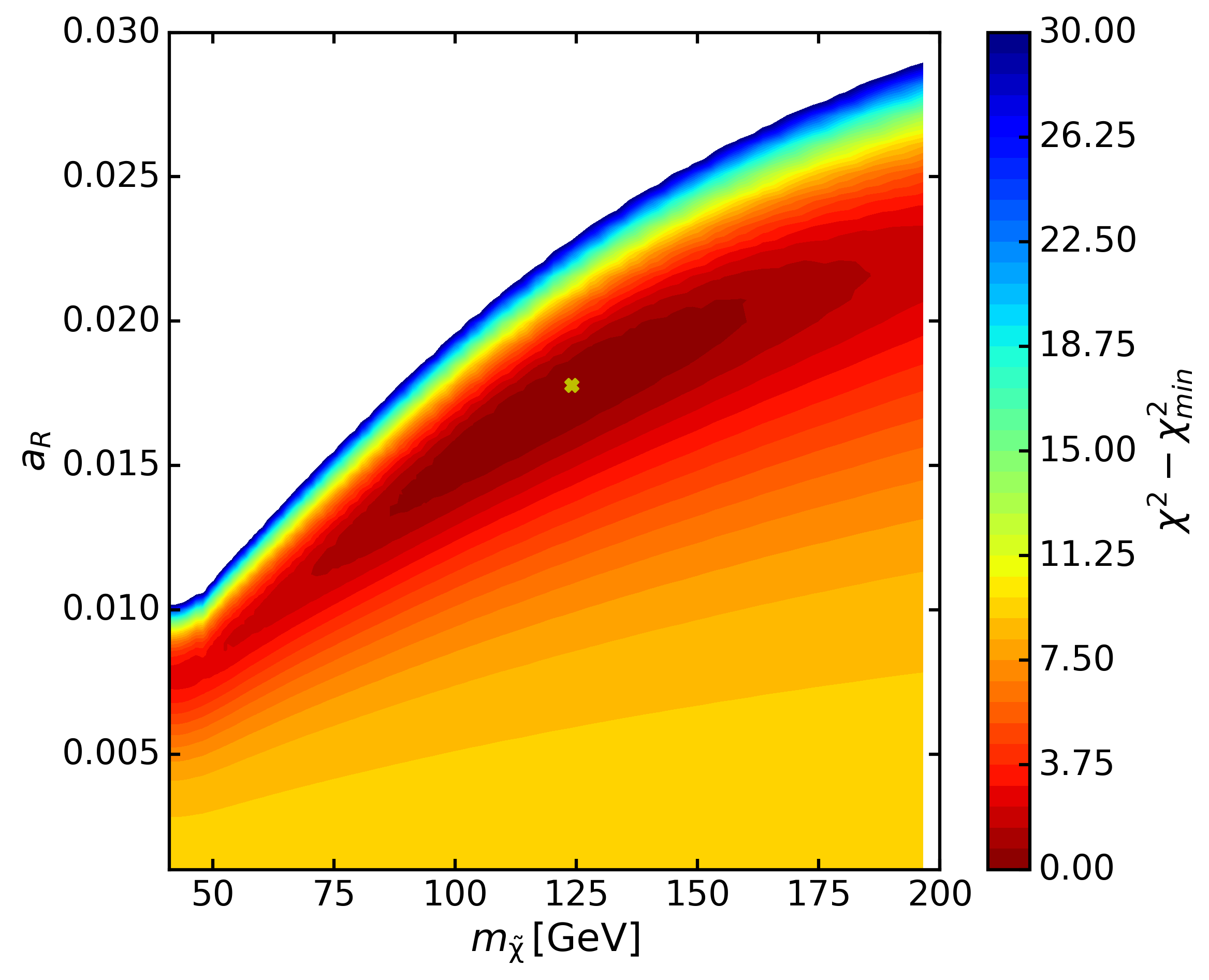}
	\includegraphics[width=0.46\linewidth]{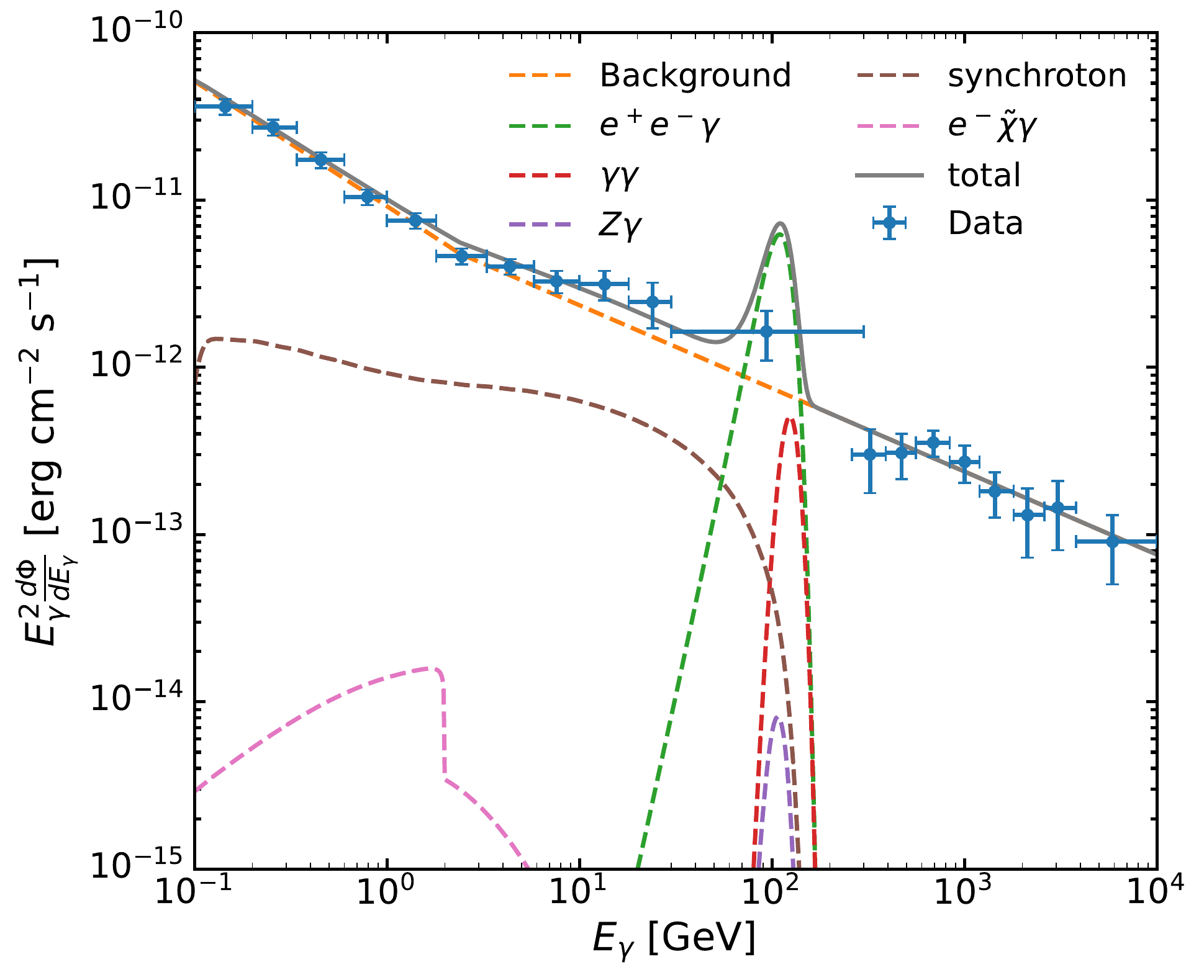}
	\caption{Left: Contour plot showing the $\chi^2$ values for the parameter space points of interest in the $m_{\tilde{\chi}}-a_R$ plane. The best fit value is indicated with a green cross. Right: Flux of the best fit parameter space point, separating the contribution of the background and of the individual channels $e^+ e^- \gamma$, $\gamma \gamma$, $Z \gamma$ and the synchroton radiation. The data points are also displayed with relative bin widths and uncertainties.}
	\label{fig:chi2}
\end{figure*}
As it is pointed out in Ref.~\cite{Abdalla:2018agf}, due to the hardening of the photon spectrum at energies above $\sim 2.5$ GeV and extending to $\sim 250$ GeV, a single-zone SSC interpretation for the overall SED of the Cen A core is disfavoured and suggests the need of a new gamma ray emitting component. The lack of variability of the gamma ray emission at both GeV and TeV energies does not yet allow disentanglement of the physical nature of this component, though a jet-related origin is possible and a simple two-zone SED model fit can mimic the observed signal.
Even if in the previous section we have considered the SSC model with two zones derived in Ref.~\cite{Abdalla:2018agf}, which is only valid if we assume that the non-existent variability measured is the result of limited statistics, the origin of the photons observed from Cen A is not completely clear. Some studies, such as Ref.~\cite{Brown:2016sbl}, suggest that the photons with energies between $2.5$ GeV and $5$ TeV can be partially explained by the self-annihilation of DM particles. The limited angular resolution of current gamma ray instruments together with the non-existent variability measured, do not make it possible to strongly exclude any of these hypotheses. \C{However, as it has been previously pointed out, in Ref.~\cite{Hess:2020tvd} it has been showed that at least part of the photon emission at $E_\gamma \gtrsim 100 \; \rm GeV $ arises on large scales, \ie from the kpc scale jet.}

In the subsequent analysis, we will assume that the data measured by Fermi-LAT and HESS can be explained only partially with known astrophysical sources and we will fit the excess with DM signals. In particular, we fit energies below $2.4$ GeV and above $300$ GeV with a broken power law
\begin{equation}
\frac{d\Phi^{back}}{dE_\gamma}=\left\{\begin{array}{ll}
k E_\gamma^{-\gamma_1} & E_\gamma \le E_{br} \\
k E_{br}^{\gamma_2 + \gamma_1} E_\gamma^{-\gamma_2} & E_\gamma > E_{br} 
\end{array}\right. 
\end{equation}
where $E_{br}=2.4$ GeV, and the remaining data with DM annihilation. The total flux expression that we fit is therefore given by
\begin{equation}
\frac{d\Phi}{dE_\gamma} = \frac{d\Phi^{back}}{dE_\gamma} + \frac{d\Phi^{ann}}{dE_\gamma} \, .
\end{equation}
Regarding the annihilation flux, we convolute the expression in Eq.~\eqref{eq:ann} with a Gaussian kernel in order to smear the signal and reproduce a result closer to the experimental measurements. In particular, we have
\begin{equation}
\frac{d\Phi^{ann}}{dE_\gamma} = \int_{0}^{\infty} dE_\gamma^\prime \, K(E_\gamma, E_\gamma^\prime) \frac{d\Phi^{ann}}{dE^\prime_\gamma}(E_\gamma^\prime) \, ,
\end{equation}
where
\begin{equation}
K(E_\gamma, E_\gamma^\prime) = \frac{1}{\sqrt{2 \pi} \sigma} e^{-\frac{(E_{\gamma} - E_{\gamma}^\prime)^2}{2\sigma^2}} \, .
\end{equation}
We naively assume a $10\%$ uncertainty at the experimental level which fixes $\sigma=0.1 E_{\gamma}^\prime$.

Since we consider parameter space points that satisfy the relic abundance constraint, one degree of freedom of our model is fixed, and therefore the flux we consider is characterised by a total of $5$ free parameters, \ie $\{k, \gamma_1, \gamma_2, a_R, m_{\tilde{\chi}}\}$. Since we are fitting $N_{data}=19$ data points taken from Ref.~\cite{Hess:2020tvd}, the number of degrees of freedom is $N_{dof} = 19 - 5 = 14$ and 
\begin{equation}
\chi^2 = \sum_{i=1}^{N_{data}} \frac{(\mathcal{O}_{meas}^i - \mathcal{O}_{th}^i)^2}{\sigma_i^2}
\end{equation}
is the usual $\chi^2$ definition with $\mathcal{O}_{meas}^i$ the measured values, $\mathcal{O}_{th}^i$ the theoretical prediction and $\sigma_i$ the uncertainty of the data.

In Fig.~\ref{fig:chi2}, on the left, we show the contour plot of the $\chi^2$ values as a function of the DM mass and the coupling, identifying the parameter space point that best fits the data. The minimum value $\chi^2_{min}=1.59$ is subtracted for convenience. For this fit, we have considered the photons coming from the annihilation of DM into $\gamma \gamma$, $\gamma Z$ and $e^- e^+ \gamma$ together with the ones coming from the synchrotron radiation of the electrons produced in the annihilation in the presence of a magnetic field. In order to obtain the latter, we consider the equipartition magnetic
field model, see Refs.~\cite{Regis:2008ij,Lacroix:2015lxa}, and therefore the magnetic field can be written as

\begin{equation}
B(r)=\left\{\begin{array}{ll}
B_{0}\left(\frac{r_{\mathrm{c}}}{r_{\mathrm{acc}}}\right)^{2}\left(\frac{r}{r_{\mathrm{acc}}}\right)^{-\frac{5}{4}} & r<r_{\mathrm{acc}} \\
B_{0}\left(\frac{r}{r_{\mathrm{c}}}\right)^{-2} & r_{\mathrm{acc}} \leq r<r_{\mathrm{c}} \\
B_{0} & r \geq r_{\mathrm{c}}.
\end{array}\right.  
\label{eq:B}
\end{equation}
$B_0 \sim 200 \; \mu \rm G$ is obtained assuming a mean comoving magnetic field of $B \sim 10$ G, $r_{\rm acc}=2 G M_{\rm BH}/ \nu_{\rm flow}^2$ is the accretion radius, with $\nu_{\rm flow} \sim 500-700 \; \rm km/s$ the velocity of the Galactic wind at the center of Cen A, and $r_c \sim 5 \; \rm kpc$ the radius of the inner cocoon, \ie the size of the inner radio lobes. Note that the mean comoving magnetic field for the jet model used in the previous section is $6.2$ G for the first SSC zone and $17$ G for the second. In this section we are using a different assumption regarding the origin of the measured flux of photons but we consider a mean magnetic field close to these values. It is important to notice too that the 1st SSC zone model (without including the 2nd zone) can be compatible with the scenario of this Section and therefore we think that a value of $B \sim 10$ G is a reasonable choice. Besides, we have verified that the dependence of the flux on the comoving magnetic field is approximately $\Phi_{\rm syn} \propto 1/B$, for values of $B$ not too far from $10$ G. However, the complete dependence on $B$ is rather complex and the shape of the flux can be affected. We checked for an envelope of values between $[5, 20]$ G, finding that the shape of the flux does not change and an overall rescaling of the synchroton flux would be sufficient to describe a modification of the magnetic field in that range.
Details on how to perform the calculation of the synchrotron radiation signal can be found in the Appendix A.

It is worthy to mention that photons coming from synchrotron radiation can be circularly polarised. However, even for high magnetic fields the circular polarisation asymmetry is expected to be very low for photons in the gamma ray band. This can be inferred from Ref.~\cite{deBurca:2015kea}, where the authors find that the circular polarisation fraction of photons coming from synchrotron radiation in high magnetic fields scales as $E_\gamma^{-p/2}$ for an electron energy spectrum $\propto E_{e}^{-p}$. In our case $p \sim 2$. Since their results show that a fraction of circular polarisation higher than $50\%$ cannot be reached by photons with energies of $\sim 2$ eV, the synchrotron radiation coming from the annihilation of our DM candidate is expected to have a negligible circular polarisation asymmetry.

Going back to the results of our fit, we observe that the best fit value in Fig.~\ref{fig:chi2} is found for a DM candidate with $m_{\tilde{\chi}}= 123$ GeV, $\Delta M = 2$ GeV and $a_R = 0.018$. The respective flux is shown in the right panel of Fig.~\ref{fig:chi2}, where all the different components considered in our calculation are displayed\footnote{Note that the flux displayed in the figure is the differential flux, while the fit is performed by computing the differential binned flux, \ie the integral of the differential flux in each bin divided by the bin width.}. However, given the large bin widths, especially of the bin $[30, 300]$ GeV, several DM candidates show a good fit. In order to better discern the various models, a finer binning would be needed.

In the right panel of Fig.~\ref{fig:chi2}, we also display the corresponding photon flux which comes from the scattering of DM and electrons in the jet and provide a circularly polarised signal. For the calculation of this flux we have considered the electron energy spectrum derived in Ref.~\cite{Abdalla:2018agf}, which assumes a two zone SSC model explaining all the photons measured with SM processes. However, for the value of the mass splitting showed, only the 1st SSC zone is relevant and therefore the treatment is consistent with the scenario of this Section. As we expected, this photon flux is very low with respect to the components coming from annihilation and therefore it is not worthy to implement it in the fit and we just show it here for comparison.

It is relevant to mention that both the DM candidate which provides the best fit value and the candidates allowed in the previous scenario considered, \ie $40\; \textrm{GeV} \lesssim m_{\tilde{\chi}}\lesssim 200$ GeV and $a_R \lesssim 10^{-2}$, will not provide observable signatures in direct detection and collider searches in upcoming years. In Figure 5 of Ref.~\cite{Cermeno:2021rtk}, constraints on the anapole moment from direct detection for this model are reported. Note that this candidate can interact with quarks only via one loop anapole moment. By looking at the figure, we see that not only our candidate is not excluded by XENON1T, but also it is not expected to be observed by the next-generation LZ experiment. With respect to collider searches,  constraints based on the pair production of the mediator and its subsequent decay from the LHC exclude only scenarios in which the mediator mass is $m_\varphi \lesssim 100$ GeV and $\Delta M \sim \textrm{few}$ GeV, while LEP can only put bounds for $m_{\varphi} \lesssim 90$ GeV and $m_{\varphi}/m_{\chi} \geq 1.03$, see Ref.~\cite{Cermeno:2021rtk} for more details on this. Furthermore, in Ref.~\cite{Liu:2021mhn} a summary of the current LHC and LEP constraints on this model based on the mediator decay and projections for the future Circular Electron Positron Collider are reported. Note however that lower mass splittings of less than a few GeV could be explored by the future $e^+$ $e^-$ colliders such as the International Linear Collider (ILC) \cite{Berggren:2013vna}. Another possible signal produced by the DM model considered is mono-photon signatures. However, as it is shown in Ref.~\cite{Kopp:2014}, both LEP and the projected ILC limits can only exclude candidates displaying annihilation cross-sections above $\sim 10^{-27}\; \rm cm^3/s$. A different future linear collider which could reach center of mass energies high enough to produce our DM candidates is the Compact Linear Collider (CLIC)~\cite{2012arXiv1202.5940L}. In order to estimate the prospects of CLIC to observe this kind of signatures, we have computed the cross section for $e^- e^+ \rightarrow \tilde{\chi} \tilde{\chi} \gamma$ at $\sqrt{s}=3$ TeV for the best fit parameter space point, finding $\sigma \sim 10^{-6}$ fb. This cross section is 6 orders of magnitude below the projected sensitivity of the future linear collider, estimated to be of few fb~\cite{Blaising:2021vhh} for mono-photon searches.
Therefore, we can conclude that the DM candidate considered could be possibly detected either by indirect detection of its annihilation and scattering products in high DM density regions or via the mediator pair production at future $e^+$ $e^-$ colliders.

\section{Conclusions}
\label{sec:conclusion}
In this work we have explored for the first time the photon signatures of leptophilic t-channel thermal DM in Cen A, the closest active galaxy powered by its AGN. This AGN, apart from being an important source of high-energy electrons, is expected to possess a DM density spike in its core due to the DM accretion onto its  SMBH, which should have survived to date. This is not the case for the Milky Way since our galaxy is old enough so that the spike structure is smoothed down due dynamical relaxation caused by the scattering of DM with stars. Therefore, the study of the signatures coming from the DM rich environment of Cen A can complement and even improve the constraints coming from conventional searches in our galaxy.

In the model we consider, where our DM candidate is a Majorana fermion which couples to right-handed electrons via a scalar mediator, both the DM self-annihilations, $\tilde{\chi} \tilde{\chi} \rightarrow e^- e^+ \gamma$, $\tilde{\chi} \tilde{\chi} \rightarrow \gamma \gamma$, and its scatterings with electrons, $\tilde{\chi} e^- \rightarrow \tilde{\chi} e^- \gamma$, can resemble line-like signatures in the photon spectrum. These peaks on the spectrum are expected at energies equal to the DM particle mass and mass splitting respectively. Moreover, since the DM-electron scattering arise via a P violating interaction, the photons produced in these interactions will be circularly polarised.

Given the fact that the origin of the X ray and gamma ray photons coming from the core of Cen A measured by Fermi LAT and HESS is not completely clear, we have considered two different scenarios in order to derive our results.
Firstly, following Ref.~\cite{Abdalla:2018agf} we assume that the photons detected in the whole energy range of the spectrum can be explained completely by the SSC radiation of two emitting zones. Consequently, considering the measured photons as a background for potential DM signals, we derive constraints over our model due to the self-annihilation of DM. In particular, DM candidates with average annihilation cross-sections below $\langle\sigma v\rangle \sim 10^{-34}-10^{-35}$ cm$^3$/s are ruled out. \C{These constraints are 7 orders of magnitude stronger than the ones from the GC} and imply that $m_{\tilde{\chi}} \lesssim 200 \; \rm GeV$ for thermal DM candidates. Besides, measurements of Z boson decay into invisible particles leave basically no room for an additional decay channel into the dark sector. In order to avoid this constraint, the mass of the mediator is restricted to  $m_{\varphi} \geq 45$ GeV and, therefore, no DM mass smaller than $\sim 40$ GeV is allowed for mass splittings of a few GeV. In summary, the allowed range of DM masses is $40 \; \textrm{GeV} \lesssim m_{\tilde{\chi}} \lesssim 200$ GeV with couplings $a_R \lesssim 10^{-2}$. In this region the mass splittings are fixed by the relic density constrain, since co-annihilations dominate and totally determine the relic, and in particular we have $\Delta M = 3 \, (1)$ GeV for a DM candidate with mass $m_{\tilde{\chi}}=50 \, (200)$ GeV. 

Focusing on the allowed area of the parameter space, we calculate the photon flux coming from the radiative scattering of DM particles with the electrons of the jet and the circular polarisation asymmetry of these photons. We found that the flux of photons coming from $\tilde{\chi} e^- \rightarrow \tilde{\chi} e^- \gamma$ in Cen A is two orders of magnitude lower than the background. The circular polarisation asymmetry related to these interactions reaches values close to 100$\%$, a degree of asymmetry which cannot be mimicked by any other SM source. Therefore, a measurement of strong polarization would be a proof of new physics. However, new techniques to measure the circular polarisation of photons should be developed if we plan to take advantage of this feature.

In the second scenario, we consider that the photons measured by Fermi LAT and HESS can be explained only partially with SM processes in the jets, \ie SSC radiation, and we fit the excess with DM signals coming from self-annihilation. The resulting best fit value is obtained for a DM candidate with $m_{\tilde{\chi}}= 123$ GeV, $\Delta M = 2$ GeV and $a_R = 0.018$.

In summary, both scenarios point to a DM candidate with a mass $m_{\tilde{\chi}} \sim 100$ GeV and couplings $a_R \lesssim 10^{-2}$. Therefore, if a DM spike is present in the inner core of Cen A as it is argued in Ref.~\cite{Gondolo:1999ef}, a leptophilic Majorana fermion interacting with electrons through a scalar mediator could be a DM candidate providing the correct relic density only under these circumstances. In that case, $100\%$ circularly polarised signals of photons are expected for energies around the value of the mass splitting.





\section*{Acknowledgments}
 We thank F. Rieger, L. Ubaldi, S. Heinemeyer and J. A. Aguilar-Saavedra for helpful discussions and comments. The work of C.D. and M.C. was funded by the F.R.S.-FNRS through the MISU convention F.6001.19. The work of L.M. is supported by the European Research Council under the European Union’s Horizon 2020 research and innovation Programme (grant agreement n.950246). This work has been partially supported by STFC consolidated grant ST/T000694/1. Computational resources have been provided by the supercomputing facilities of the Universit\'e Catholique de Louvain (CISM/UCL) and the Consortium des \'Equipements de Calcul Intensif en F\'ed\'eration Wallonie Bruxelles (C\'ECI) funded by the Fond de la Recherche Scientifique de Belgique (F.R.S.-FNRS) under convention 2.5020.11 and by the Walloon Region.

\appendix
\section{Synchrotron radiation flux}
In order to compute the photon flux coming from the synchrotron radiation of electrons/positrons produced by DM annihilation we follow \cite{Lacroix:2015lxa}. For this calculation we need to perform the following integrals over the electron energy $E$ and the radial coordinate $r$
\begin{equation}
E_\gamma^2 \frac{d\Phi_{\rm syn}}{dE_\gamma} = \frac{8 \pi E_\gamma}{d_{\rm AGN}^2} \int_{4 R_S}^{r_0} r^2 dr \int_{E_\gamma}^{m_{\chi}} dE P(r, E, E_\gamma) \psi_{\mathrm{e}}(r, E), 
\end{equation}
where $\psi_{\mathrm{e}}(r, E)$ is the electron and positron energy spectrum coming from $\tilde{\chi} \tilde{\chi} \rightarrow e^- e^+ \gamma$ and $P(r, E, E_\gamma)$ is the synchrotron emission spectrum.
Specifically, for our model the electron and positron energy spectrum can be written as
\begin{equation}
\psi_{\mathrm{e}}(r, E)=\frac{1}{2 b(r, E)}\left(\frac{\rho_{DM}(r)}{m_{\mathrm{DM}}}\right)^{2} \int_{E}^{m_{\mathrm{DM}}} \mathrm{d} E_{\mathrm{S}} \frac{d \sigma v_{e^- e^+ \gamma}}{dE_S},     
\end{equation}
with $\frac{d \sigma v_{e^- e^+ \gamma}}{dE_S}$ the differential velocity averaged cross section for the annihilation channel $\tilde{\chi} \tilde{\chi} \rightarrow e^- e^+ \gamma$, which is a function of $E_S$, and 
\begin{equation}
b(r, E)=\frac{4}{3} \sigma_{\mathrm{T}} \frac{B(r)^{2}}{2 \mu_{0}} \gamma_{\mathrm{L}}^{2}    
\end{equation}
the total energy loss rate, which depends on the magnetic field $B(r)$, showed in Eq.\eqref{eq:B}, the electron Lorentz factor $\gamma_L=\frac{E}{m_e}$, the vacuum permeability $\mu_0$ and the Thomson cross-section $\sigma_T$.

The synchrotron emission spectrum can be expressed as
\begin{equation}
 P(r, E, E_\gamma)=\frac{1}{4 \pi \varepsilon_{0}} \frac{\sqrt{3} e^{3} B(r)}{m_{\mathrm{e}}} G_{\mathrm{i}}\left(\frac{E_\gamma}{E^{\mathrm{c}}_{\gamma}(r, E)}\right)   
\end{equation}
where $e$ is the elementary charge, $\varepsilon_{0}$ the vacuum permittivity, and $G_{\mathrm{i}}$ the isotropic synchrotron spectrum, which depends on the critical photon energy 
\begin{equation}
E^{\mathrm{c}}_{\gamma}(r, E)=\frac{3 e E^{2} B(r)}{4 \pi m_{\mathrm{e}}^{3}}.    
\end{equation}

$G_{\mathrm{i}}(x)$ can be obtained by averaging the synchrotron spectrum over an isotropic distribution of pitch angles 
\begin{equation}
G_{\mathrm{i}}(x)=\frac{1}{2} \int_{0}^{\pi} G\left(\frac{x}{\sin \alpha}\right) \sin ^{2} \alpha \mathrm{d} \alpha,    
\end{equation}
with $G(t)=t \int_{t}^{\infty} K_{5 / 3}(u) \mathrm{d} u$, where $K_{5 / 3}$ is the modified Bessel function of order $5 / 3$. However, in order to simplify the numerical calculation we use the parametrization provided in \cite{crusius}.



\bibliography{refs}


\end{document}